# A Suggestion for MOND v2

Peter Rastall
*Department of Physics and Astronomy*
*University of British Columbia*
*Vancouver B.C. V6T1Z1, Canada*
email: peras@shaw.ca

Modified Newtonian dynamics (MOND) has had considerable success in describing motions in galaxies. It uses a single force which falls off inversely with the distance at large distances and inversely with the square of the distance at smaller distances. We present an alternative theory with two forces, one the traditional Newtonian inverse-square, and the other that falls off inversely with distance at large distances. This obvious possibility has been avoided because of fear that the second force would be incompatible with observed planetary motions. However, the non-linear field equation that governs this force is shown to reduce its strength near stars. The theory is derived from a Lagrangian density with two scalar potentials. It is non-relativistic, but nevertheless agrees with the 'classical tests of relativistic gravity' and can be used to calculate the bending of light in gravitational fields. Possible applications to interactions in galactic clusters and to anomalous planetary motions are noted.

## 1 Introduction

It has long been known that the motions in galaxies cannot be explained by the Newtonian gravitational attraction of the visible matter [Oort 32], [Zwicky 33]. To account for the missing force, several kinds of dark matter have been postulated, but nothing has been found in sufficient quantity. Another possibility is that the Newtonian law should be changed. One must of course take care that a new law does not significantly perturb motions in the solar system, which have been accurately measured.

Mordehai Milgrom discovered in the early 1980s that the motion of stars in the outer regions of galaxies can be explained if the gravitational force falls off inversely with distance, rather than inversely with the square of the distance as in the Newtonian theory [Milgrom 83,14]. He postulated that the force becomes inverse-square at smaller distances.

In what follows, it is assumed that two forces act: the usual Newtonian inverse-square and another that behaves like Milgrom's at large distances. For simplicity, one considers a spherically symmetric galaxy of mass M and a star of mass m that moves outside this galaxy with speed v in a circular orbit of radius r. The gravitational force **f** that acts on the star is directed towards the centre of the galaxy and has magnitude f,

$$f = mMGr^{-2} + mAr^{-1}, \qquad (1.1)$$

where $G = 6.67 \times 10^{-11}$ m$^3$s$^{-2}$kg$^{-1}$ is the Newtonian gravitational constant and A is a constant that depends on M.

The acceleration **a** of the star has magnitude $v^2 r^{-1} = MGr^{-2} + Ar^{-1}$. For large r, the $r^{-2}$ term is negligible and $v^2 = A$. Hence v is constant, independent of r (one says that the galactic rotation curve is *flat*). It is found that for a large number of galaxies $v = (GMa_0)^{1/4}$, where $a_0 = (1.2 \pm 0.2) \times 10^{-10}$ ms$^{-2}$ according to [Milgrom 14], and hence

$$A = (GMa_0)^{1/2} = KM^{1/2}, \tag{1.2}$$

where $K = (Ga_0)^{1/2} = 8.9 \times 10^{-11} \, m^2 s^{-2} kg^{-1/2}$. Note that K, like $a_0$, is the same for all the galaxies. (If $c = 3 \times 10^8 \, ms^{-1}$ is the speed of light, then $K^{-2}c^4$ has the dimensions of mass and is about $10^{54}$ kg – which is roughly the mass of the observable universe.)

We recall that in traditional MOND the gravitational attraction a that is produced by a point mass M at a distance r is given by $\mu(a/a_0)a = GMr^{-2}$. Here $a_0$ is the constant that appears in (1.2) and $\mu$ is a function such that $\mu(x)$ approaches 1 for large x and $\mu(x)$ approaches x for small x. This is a radical change from the principles of Newtonian mechanics. The purpose of the present paper is to show that the introduction of the function $\mu$ is unnecessary and that standard Newtonian mechanics can be preserved – but at the cost of introducing an additional force.

The force (1.1) on the star can be written as $\mathbf{f} = -m\nabla\Phi$, where

$$\Phi = -MGr^{-1} + A \ln r. \tag{1.3}$$

Since the Laplacian of the first term vanishes, one has $\nabla^2\Phi = Ar^{-2}$, valid for the case of spherical symmetry and where the mass density of the sources vanishes. In the next Section, a field equation without these restrictions will be found.

## 2 Field equations

The potential $\Phi$ is written as $\Phi = \Theta + \Psi$ and the Lagrangian density for $\Theta$ and $\Psi$ is

$$L = \Theta_{,i} \Theta_{,i} + 8\pi G\rho\Theta + (\Psi_{,i} \Psi_{,i})^{3/2} + 6\alpha\Psi \tag{2.1}$$

where $\alpha$ is a constant, $\rho$ is the mass density of the sources of the fields, $\Theta_{,i} = \partial\Theta/\partial x_i$, etc., the $x_i$ are rectangular cartesian coordinates, and the summation convention applies to repeated latin suffixes with the range $\{1, 2, 3\}$.

The field equations that follow from (2.1) are

$$\Theta_{,i\,i} = 4\pi G\rho, \tag{2.2}$$

$$\Psi_{,j} \Psi_{,j} \Psi_{,ii} + \Psi_{,i} \Psi_{,j} \Psi_{,ij} = 2\alpha\rho (\Psi_{,i} \Psi_{,i})^{1/2}. \tag{2.3}$$

Eq.(2.2) is the usual Poisson equation for the Newtonian potential $\Theta$. In vector notation, (2.3) is

$$|\nabla\Psi|^2 \nabla^2\Psi + \tfrac{1}{2}\nabla(|\nabla\Psi|^2) \cdot \nabla\Psi = 2\alpha\rho |\nabla\Psi|, \tag{2.4}$$

or equivalently,

$$\mathrm{div}(\,|\nabla\Psi|\nabla\Psi) = 2\alpha\rho. \tag{2.5}$$

The Lagrangian density (2.1) is similar to that used in [Bekenstein 88] for the MOND theory, but without the $\mu$ terms.

In the special case of spherical symmetry, eq.(2.4) becomes

$$\Psi'' + r^{-1}\Psi' = \alpha\rho|\Psi'|^{-1}, \tag{2.6}$$

where $\Psi' = d\Psi/dr$, etc. If $\rho = 0$, (2.6) gives $\Psi' = Ar^{-1}$ with A constant, in agreement with (1.3).

One must show that A is proportional to the square root of the mass of the source. Applying the divergence theorem to a ball of radius R centred at the origin and assuming that $r^2\rho(r)$ vanishes as $r\to 0$ gives

$$|\Psi'(r)|\Psi'(R) = (\alpha/2\pi)R^{-2}I(R), \qquad (2.7)$$

from (2.5), where $I(R) = \int 4\pi\rho r^2 dr$, with integration from 0 to R, is the total mass in the ball of radius R. If $\rho(r) = 0$ for $r>R$, then $I(R) = M$, the total mass of the sources. If $\Psi'$ is assumed to be non-negative, $\Psi'(R) = (\alpha/2\pi)^{1/2} R^{-1}M^{1/2}$. When $r>R$, one has $\Psi' = Ar^{-1}$ and hence

$$A = (\alpha/2\pi)^{1/2}M^{1/2} = KM^{1/2}, \qquad (2.8)$$

where $K = (Ga_0)^{1/2}$, as in (1.2), to agree with the observations of galactic motion, and $\alpha = 2\pi K^2 = 5.03\times 10^{-20}\,m^4 s^{-4} kg^{-1}$.

Instead of $6\alpha\Psi$ in (2.1), one might have chosen another function of $\Psi$. However, this spoils the simple $M^{1/2}$ dependence of (2.8).

## 3 Near the Sun

The field equations are assumed to be generally valid. They can therefore be used to find the effect of the $\Psi$ field on planetary motions. For simplicity, we assume a spherically symmetric Sun of mass $M = 2\times 10^{30}\,kg$. The force per unit mass exerted by the $\Psi$ field at a distance r from the Sun's centre has magnitude $\Psi'(r) = KM^{1/2}r^{-1}$ and that exerted by the $\Theta$ field has magnitude $\Theta'(r) = GMr^{-2}$. At the orbit of Mercury, radius $5.8\times 10^{10}\,m$, one has $\Psi'(r) = 2.17\times 10^{-6}\,ms^{-2}$ and the ratio of these forces is $KR/GM^{1/2} = 5.5\times 10^{-5}$. This is to be compared with $GM/rc^2 = 2.6\times 10^{-8}$, which is roughly the ratio of the post-Newtonian perturbation force to the Newtonian force exerted by the $\Theta$ field. The force exerted by the $\Psi$ field is therefore three orders of magnitude greater than the observed perturbation force on Mercury.

The conclusion is clear: the $\Psi$ field is not in agreement with the observed planetary motions. This was of course obvious to Milgrom, and is why he chose a force field that did not have an $r^{-1}$ dependence at small distances.

But this cannot be the whole story. For simplicity again, suppose that the galaxy has $10^{11}$ stars of mass M, distributed in a spherically symmetric manner. According to our previous result, each of them produces a $\Psi$ field proportional to $M^{1/2}$, and the total $\Psi$ field of the galaxy might therefore be expected to be $10^{11/2}$ as great $(10^{11} M^{1/2} = 10^{11/2} (10^{11} M)^{1/2})$. But we know from eq.(2.8) that the correct factor is 1, not $10^{11/2}$. The assumption that one can add together the contributions of the individual stars is incorrect – what prevents it is the non-linearity of the field equations. We must show how the fields of the stars are melded together to give the total field of the galaxy. We may expect that this will diminish the fields near the individual stars by large factors.

So much for hand-waving! To show how the $\Psi$ field is repressed, one considers a large number of stars uniformly distributed in a thin, spherical shell of radius R and mass $M_S$. One makes the usual assumption that the stars can be represented by a fluid of density $\rho$. Since $\rho(r)$ and $I(r)$ vanish inside the sphere, it follows from (2.7) that so does $\Psi'(r)$ (we are assuming that $\Psi$ and $\Psi'$ are non-singular inside the shell). Outside the shell, one finds as in (2.8) that $\Psi' = Ar^{-1}$, where $A = KM_S^{1/2}$. If $M_S = 10^{38}\,kg$ (about $0.5\times 10^8$ solar masses) and $R = 3.5\times 10^4$ ly = $3.3\times 10^{20}\,m$ (roughly the Sun's distance from the centre of the galaxy), one has $\Psi'(R) = 2.7\times 10^{-12}\,ms^{-2}$. Compare this with $2.17\times 10^{-6}\,ms^{-2}$, the naively calculated value of $\Psi'$ due to the Sun at the orbit of Mercury. If we now abandon the fluid description and consider the stars in a small part of the shell, this means that they produce no $\Psi$ field towards the inside of the shell and that the field that they produce towards the outside depends on the total mass of the shell, and is six orders of magnitude less than that found by the naive calculation.

The last paragraph considered a thin, spherical shell of stars. To understand better how the $\Psi$ field of an individual star is repressed, one should solve the $\Psi$ field equation for the galaxy and an individual star, both idealized as spherically symmetric balls of fluid. (One can consider the two cases of the star inside or outside the galaxy.) Using these solutions, one can check whether the force on the star is $-m\nabla\Phi_G$, where $\Phi_G = \Theta_G + \Psi_G$ is the potential due to the galaxy. The essential condition is probably that the mass of the star is very small compared with the mass of the galaxy.

An exactly similar calculation would serve to calculate the $\Psi$ field of two galaxies, each idealized as a spherically symmetric ball of fluid. It seems unlikely that the force on one of them can be expressed in terms of the fields produced by the other – and indeed, for a non-linear field this statement makes no sense. We note that traditional MOND does not predict the correct forces between galaxies: perhaps the present approach will be more successful. The completion of these ideas is left as an exercise for the reader (the author is now too frail to handle non-linear partial differential equations).

The $\Psi$ field of the Sun will not be completely repressed. It will be interesting to see if it accounts for the remaining small anomalies in solar system motions [Anderson 09].

## 4  Gravitational calculations

The search for dark matter has so far been fruitless. It appears likely that the explanation of galactic motions will require a new gravitational force law and the abandonment or modification of the Newtonian and Einsteinian theories. This does not however imply that we shall lose the ability to calculate gravitational effects, at least in the non-relativistic domain. It has been known for many years that the Newtonian theory of gravitation, with an almost trivial generalization, can account for all the so-called 'classical tests of relativistic gravity'. These are the gravitational redshift, the anomalous perihelion advance of Mercury, and the deflection of light in a gravitational field (and one might add the radar time delay) [Rastall 79, 60]. An outline of this generalized Newtonian theory, which has been called *gravitostatics*, is in Section 1 of [Rastall 04]; a longer account is in [Rastall 91].

The theory described in this paper is easily incorporated into gravitostatics: one simply interprets the potential $\Phi$ as $\Theta + \Psi$, as in Section 2. Chapter 3 of [Rastall 91] gives a Lagrangian for the motion of particles, including photons, which enables the calculation of light deflection and gravitational lensing.

## 5  Conclusion

It is not necessary to modify Newtonian dynamics in a radical way, as originally proposed in MOND. One has only to introduce an additional force which is produced by a non-linear field. The non-linearity represses the short-range effects of this force on planetary motions.

The theory proposed is non-relativistic. It is sufficient for dealing with galactic motions, gravitational lensing, etc. It would not be difficult to produce a relativistic generalization (cf. [Bekenstein 11]), but this may be premature. It may be advisable to first explore further the properties of the additional force, and to try to account for the motion of galaxies in clusters.